\documentclass[prl,aps,twocolumn]{revtex4}

\usepackage{epsfig}

\begin{document}
%\title{Hidden scale invariance in van der Waals liquids: A simulation study of two simple molecular liquids}
\title{Hidden scale invariance in molecular van der Waals liquids: A simulation study}
%\title{Density scaling in strongly correlating viscous liquids}
\author{Thomas B. Schr{\o}der,  Ulf R. Pedersen, Nicholas Bailey, S\o{}ren Toxvaerd, and Jeppe C. Dyre}
\affiliation{DNRF Centre ``Glass and Time,'' IMFUFA, Department of Sciences, Roskilde University, Postbox 260, DK-4000 Roskilde, Denmark}
\date{\today}

\begin{abstract}
%There is a "hidden" approximate scale invariance in van der Waals liquids. 
Results from molecular dynamics simulations of two viscous molecular model liquids --  the Lewis-Wahnstr\"om model of ortho-terphenyl and an asymmetric dumbbell model -- are reported. We demonstrate that the liquids have a ``hidden'' approximate scale invariance: Equilibrium potential energy fluctuations are accurately described by inverse power law (IPL) potentials, the radial distribution functions are accurately reproduced by the IPL's, and the radial distribution functions obey the IPL predicted scaling properties to a good approximation. IPL scaling of the dynamics also applies -- with the scaling exponent predicted by the equilibrium fluctuations. In contrast, the equation of state does not obey the IPL scaling. We argue that our results are general for van der Waals liquids, but do not  apply, e.g., for hydrogen-bonded liquids.
\end{abstract}
\maketitle

\section{Introduction}

A phenomenon is scale invariant if it has no characteristic length or time. Scale invariance emerged as a paradigm in the early 1970's following the tremendous successes of the theory of critical phenomena. Liquids described by scale-invariant potentials are a theorist's dream by having a number of simple properties \cite{IPL}. If potentials vary with distance as $r^{-n}$, the configurational free energy $F$ as a function of density $\rho\equiv N/V$ and temperature $T$ may be written $F = Nk_BTf(\rho^\gamma/T)$ where $\gamma=n/3$. In fact, all properties of inverse power law (IPL) liquids -- in appropriately defined reduced units -- are functions of $\rho^\gamma/T$ only. Even dynamical properties are simple, e.g., the relaxation time $\tau$ may be written as
\begin{equation}\label{1}
\tau = t_0 \,g(\rho^\gamma/T) 
\end{equation}
with the same exponent $\gamma$ that characterizes the thermodynamics, $t_0$ being the reduced time unit ($t_0 \equiv \rho^{-1/3}\sqrt{m/k_BT}$). Unfortunately the predicted equation of state does not fit data for real fluids, and IPL liquids do not have low-pressure liquid states. Both problems derive from the absence of molecular attractions; these define the spatial scale of one intermolecular distance at low and moderate pressure. Thus real liquids have apparently little in common with scale-invariant liquids. In this paper we demonstrate by example that there is nevertheless a ``hidden'' approximate scale invariance in a large class of liquids, the van der Waals liquids. Consequently several properties of IPL liquids apply to van der Waals liquids, leading to a number of experimentally testable predictions.

\section{Strongly correlating liquids}

It was recently shown  \cite{ped_pre,ped_prl,nick} that several model liquids exhibit strong correlations between the equilibrium fluctuations of the potential energy $U$ and the virial $W$ (defining the contribution to pressure coming from the intermolecular interactions via $pV=Nk_BT+W$). ``Strongly correlating liquids'' %\cite{ped_prl, nick}  
include \cite{nick} the standard Lennard-Jones liquid, a liquid with exponential short-range repulsion, the Kob-Andersen binary Lennard-Jones liquid, a seven-site united-atom model of toluene, 
%and the two molecular models studied in detail below; 
the three-site Lewis-Wahnstr\"om  model of ortho-terphenyl (OTP), as well as a model consisting of asymmetric ``dumbbell'' type molecules.

For inverse power-law (IPL) potentials $\propto r^{-n}$ there is 100\% correlation between virial and potential energy fluctuations as function of time: $\Delta W(t)=\gamma\Delta U(t)$ where $\gamma=n/3$ \cite{IPL}. By reference to the  single-component Lennard-Jones system it was argued in Ref. \cite{nick} that strongly correlating liquids are well approximated by inverse power-law potentials as regards their thermal equilibrium fluctuations. The present paper presents concrete proofs that this is the case -- even for {\it molecular} model liquids with van der Waals type interactions (i.e., excluding hydrogen, ionic, and covalent bonds). The main conclusions below are: 

\begin{itemize}
\item {\it At a given density and temperature the energy surface of a strongly correlating molecular liquid may be replaced to a good approximation by that of inverse power-law potentials.}
\item {\it Strongly correlating liquids ``inherit'' a number of scaling properties from inverse power-law potentials, but not all such such properties.}
\end{itemize}

\section{Fluctuations and structure}

\begin{figure}
\begin{center} 
 \includegraphics[width=8.5cm]{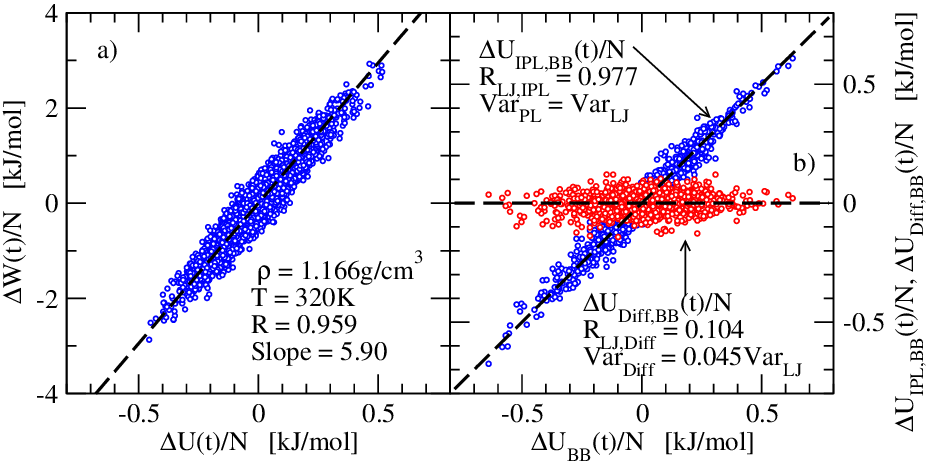}
 %\includegraphics[width=12.5cm]{fig1.eps}
 % FsB.eps: 1197x34 pixel, 0dpi, -3674644960619337680330868930316861440.00x-104375884446234152441730942875205632.00 cm, bb=
\end{center}
\caption{(Color online) 
(a) $\Delta W(t)/N$ vs.  $\Delta U(t)/N$ for the asymmetric dumbbell model \cite{Dumbbell}. The strong correlations are quantified by the correlation coefficient, $R\equiv \left< \Delta W \Delta U\right>/\sqrt{ \left< (\Delta W)^2 \right>\left< (\Delta U)^2 \right>}=0.959$. The dashed line has the slope $\gamma \equiv \sqrt{\left< (\Delta W)^2 \right>/\left< (\Delta U)^2 \right>}=5.90$. 
(b) For configurations taken from the simulation reported in Fig.~\ref{Fig:Fluct}(a) we evaluated 
%$U_{PL,\alpha \beta}(t) \equiv \sum_{i>j} \phi_{IPL,\alpha \beta}(r_{ij}(t))$ 
%with $\phi_{IPL,\alpha \beta}(r) = 4C_{\alpha \beta}\epsilon_{\alpha \beta}(\sigma_{\alpha \beta}/r)^{18}/3$
$U_{\rm IPL,BB}(t) \equiv \sum_{i>j} \phi_{\rm IPL,BB}(r_{ij}(t))$ 
with $\phi_{\rm IPL,BB}(r) \equiv C_{\rm BB}\epsilon_{\rm BB}(\sigma_{\rm BB}/r)^{18}$. The constant $C_{\rm BB}=1.489$ ($C_{\rm AA}=1.075$, $C_{\rm AB}=1.237$) was chosen to make the variance of $U_{\rm IPL,BB}(t)$ equal to the variance of $U_{\rm BB}(t)\equiv \sum_{i>j} \phi_{\rm LJ,BB}(r_{ij}(t))$. 
%(i.e., the actual total potential energy of the BB interactions). 
$U_{\rm IPL,BB}(t)$ strongly correlates with $U_{\rm BB}(t)$ with a correlation coefficient of 0.977. In comparison $U_{\rm Diff,BB}(t)\equiv U_{\rm BB}(t)-U_{\rm IPL,BB}(t)$ has small variance, $0.045$ times the variance of $U_{\rm BB}(t)$ and is only weakly correlated with the latter. %(R=0.104).
Redoing the analysis with $n=3\times 5.90$ (the IPL exponent suggested by the fluctuations, see Fig.~\ref{Fig:Fluct}(a) gives almost identical results.
}\label{Fig:Fluct}
\end{figure}

We performed NVT molecular dynamics simulations \cite{SimDet,gromacs,NoseHoover,LINCS} of two molecular liquids: i) 512 asymmetric dumbbell molecules consisting of pairs of different sized Lennard-Jones (LJ) spheres connected by rigid bonds; the dumbbells were parameterized to mimic toluene \cite{Dumbbell}; ii) the Lewis-Wahnstr\"om OTP model \cite{OTP} (N=324). Both models are strongly correlating for the range of state points investigated here. 
%(Except for pure IPL models, the ``strongly correlating''-property holds only for a region of state points \cite{nick}). 
For the dumbbell model we find for the correlation coefficient $0.95<R<0.97$; for the OTP model $0.91<R<0.92$. Fig.~\ref{Fig:Fluct}(a) illustrates the strong WU-correlation for the asymmetric dumbbell model. 
%We define $\gamma \equiv \sqrt{\left< (\Delta W)^2 \right>/\left< (\Delta U)^2 \right>}$. 
Since $R\approx 1$ it follows that $\Delta W(t) \approx \gamma \Delta U(t)$ in their instantaneous fluctuations \cite{ped_prl}, consequently $\gamma$ is referred to as the ``slope.''

Fig.~\ref{Fig:Fluct}(b) details the origin of the strong correlations, focusing on the strongest interaction in the dumbbell model, the 'BB' interaction ('B' indicating the large ``phenyl'' spheres). The fluctuations in $U_{\rm BB}(t)$ are well described by an IPL pair potential $\phi_{\rm IPL,BB}(r) \propto r^{-18}$. The same conclusion applies for the AA and AB interactions (data not shown). The reason that $\phi_{\rm Diff,\alpha\beta}(r) \equiv \phi_{\rm LJ,\alpha\beta}(r) - \phi_{\rm IPL,\alpha\beta}(r)$ contributes little to the fluctuations at constant volume is that this term is approximately linear in the region of the first peak of the pair correlation function. This implies that, when one pair-distance is increased, other pair-distance(s) decrease by approximately the same amount, keeping the change in $U_{\rm Diff,\alpha\beta}(t) \equiv \sum_{i>j} \phi_{\rm Diff,\alpha\beta}(r_{ij}(t))$ small  \cite{nick}. Switching the ensemble from NVT to NpT (p=1.32GPa) reduces the WU-correlation from 0.959 to 0.866, illustrating the importance of the constant volume constraint % in the argument above.% 
(the importance increases with decreasing pressure). 

\begin{figure}
\begin{center} 
 \includegraphics[width=8.5cm]{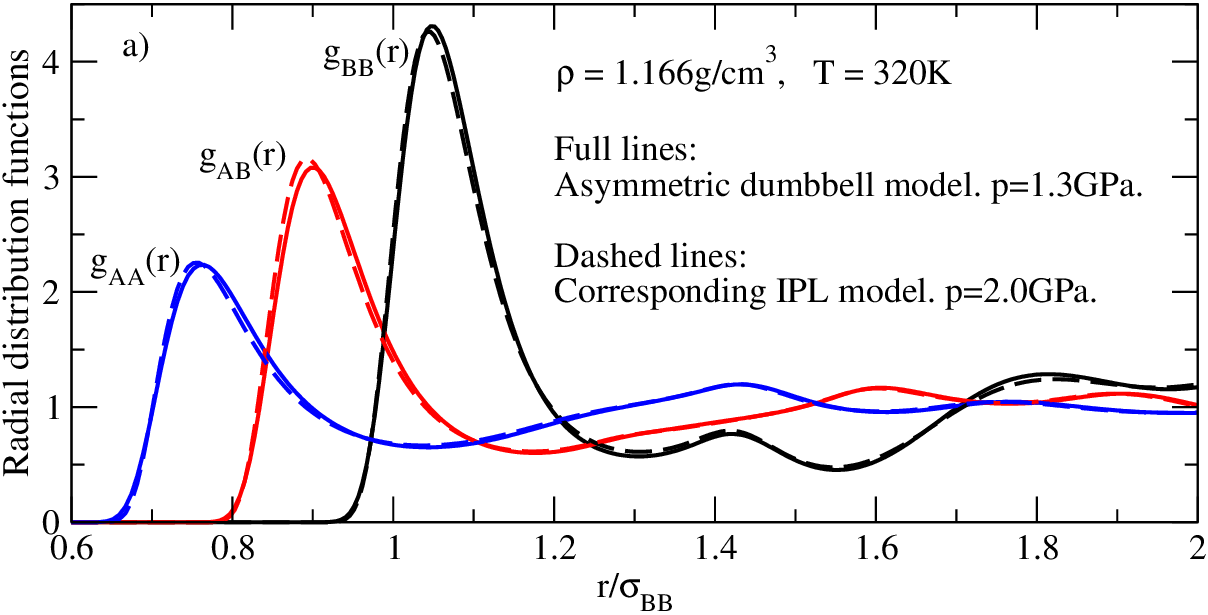}
 \includegraphics[width=8.5cm]{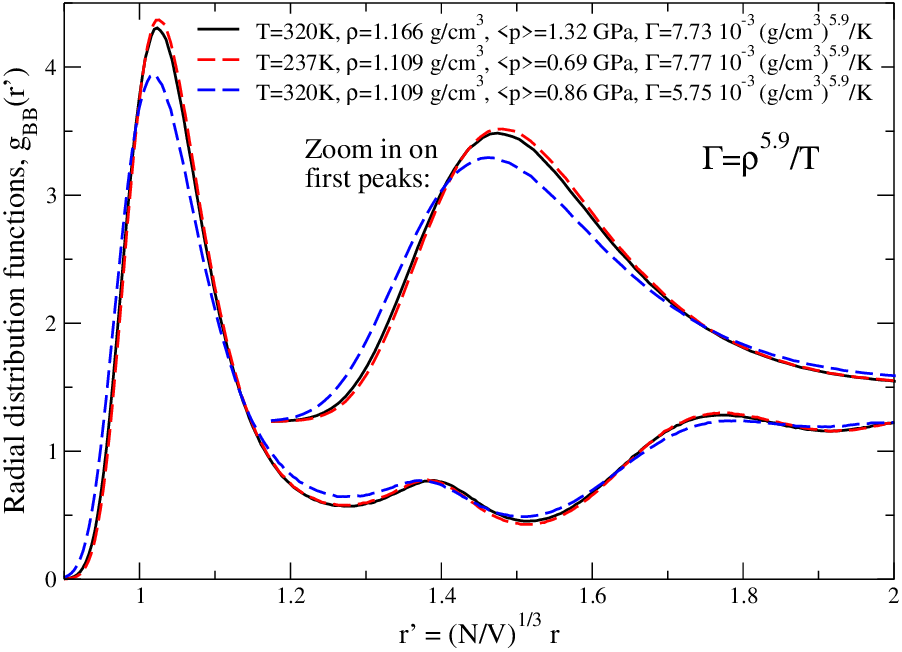}
 %\includegraphics[width=12.5cm]{fig2a.eps}
 %\includegraphics[width=12.5cm]{fig2b.eps}
 % FsB.eps: 1197x34 pixel, 0dpi, -3674644960619337680330868930316861440.00x-104375884446234152441730942875205632.00 cm, bb=
\end{center}
\caption{(Color online)
(a) Radial distribution functions for the asymmetric dumbbell model \cite{Dumbbell} (full lines), and the corresponding $r^{-18}$ IPL model (dashed lines, see text).  (b) Scaled $g_{BB}(r)$ for three state points; the two  state points with the same $\Gamma=\rho^{5.9}/T$ (black full line and red long dashed line) have almost identical (scaled) structure. When increasing $\rho$ at constant $\Gamma$ there is a slight shift of the first peak to smaller distances, consistent with the existence of the fixed bond. Similar scaling is found for the AA and AB interactions (data not shown).
%, but with slightly larger differences at short distances consistent with the fact that these interactions are weaker than the BB interaction.  
}\label{Fig:gr}
\end{figure}

The fluctuations tell us that the relevant part of the potential energy surface is well approximated by a corresponding 'IPL' system, i.e., the system where the LJ pair potentials are replaced by the IPL potentials discussed in Fig.~\ref{Fig:Fluct}(b). To test how far this correspondence holds, we simulated the IPL system at the same density and temperature. Fig.~\ref{Fig:gr}(a) compares the pair distribution functions. The agreement is striking. To the best of our knowledge this is the first proof that the structure of a molecular liquid is well reproduced by an IPL liquid. In contrast, the average pressures of the two models are quite different (see Fig.~\ref{Fig:Fluct}(b)), because the IPL system does not include contributions to the pressure from $\phi_{\rm Diff,\alpha\beta}(r)$.

\begin{figure}
\begin{center} 
 \includegraphics[width=8.5cm]{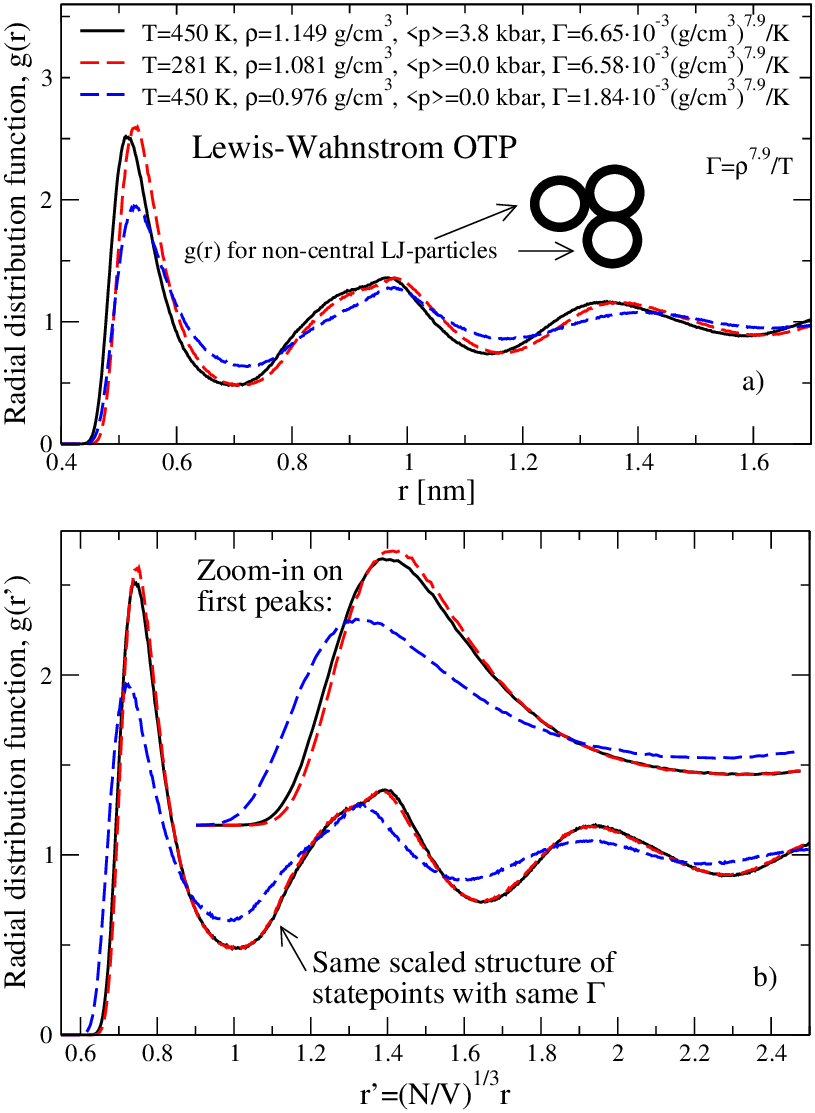}
\end{center}
\caption{(Color online)
(a) Radial distribution functions for the LW-OTP model \cite{OTP} at three different state points. Two of the state points (full black lines and red dashed lines) have approximately the same $\Gamma=\rho^{7.9}/T$. The third state point (full green line) have the same temperature as the first state point and same pressure as the same state point. 
(b) Same as in (a), but in reduced units. The two  state points with the same $\Gamma$ (black full line and red long dashed line) have almost identical (scaled) structure. Similar to what is seen for the asymmetric model (Fig.~\ref{Fig:gr}), there is a slight indication of the fixed bond in the first peak of the two state points with the same $\Gamma$ 
}\label{Fig:grOTP}
\end{figure}

Consider now changing density and temperature, conserving the parameter $\Gamma=\rho^{\gamma}/T$. For a pure  $n=3\gamma$ IPL liquid this means that one is effectively studying the same system, but with scaled length and time units (times scaled by $t_0$, lengths by $l_0\equiv \rho^{-1/3}$) \cite{IPL}: Two state points with the same $\Gamma$ have 3N-dimensional potential energy surfaces that are identical in reduced units. Switching back to the LJ system amounts to adding the ``diff'' terms of the potentials, which -- to a good approximation -- simply shifts the whole energy surface by an amount that only depends on the volume of the state point \cite{nick}. This changes neither structure nor dynamics, and consequently these properties are predicted to ``inherit'' the scaling properties of the IPL potentials. In contrast, the shift of the energy surface does change the absolute values of potential energy and pressure (because the shift depends on volume), and these properties are \emph{not} predicted to inherit the IPL scaling. 

\begin{figure}
\begin{center} 
 \includegraphics[width=8.cm]{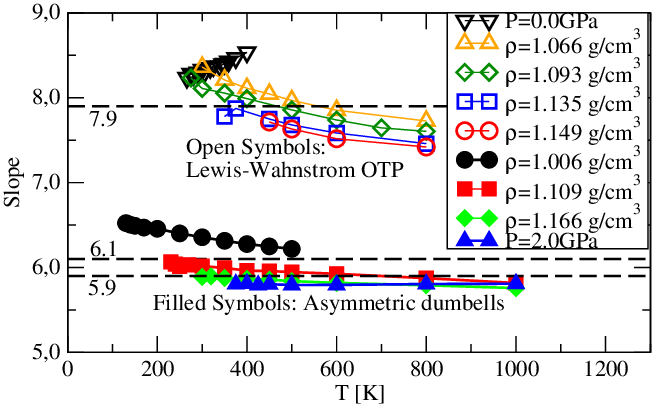}
 %\includegraphics[width=12.cm]{fig4.eps}
 % FsB.eps: 1197x34 pixel, 0dpi, -3674644960619337680330868930316861440.00x-104375884446234152441730942875205632.00 cm, bb=
\end{center}
\caption{(Color online)
%Results from equilibrium molecular dynamics simulations of two molecular liquids; an asymmetric dumbbell model (filled symbols), and the Lewis-Wahnstr\"om OTP model (open symbols). Both models are strongly correlating. 
The 'slopes', $\gamma \equiv \sqrt{\left< (\Delta W)^2 \right>/\left< (\Delta U)^2 \right>}$ plotted as a function of temperature for two molecular models. Both models are strongly correlating. For the Lewis-Wahnstr\"om OTP model \cite{OTP} the zero pressure isobar covers densities in the interval 1.008 g/cm$^3$ to  1.089 g/cm$^3$. For the  asymmetric dumbbell model \cite{Dumbbell} the 2.0GPa isobar covers densities in the interval 1.120 g/cm$^3$ to  1.217 g/cm$^3$. For the three dumbbell isochores the pressures at the lowest temperatures are -0.02GPa, 0.67GPa, and 1.28GPa respectively.
}\label{Fig:Slope}
\end{figure}

The IPL scaling of the structure is tested in Fig.~\ref{Fig:gr}(b) for the asymmetric dumbbell model, and in Fig.~\ref{Fig:grOTP} for the LW-OTP model. For both models the IPL scaling works well; there is only a small effect of the fixed bond connecting the LJ spheres. Note that average pressure and potential energy do \emph{not} follow the IPL scaling.

\section{Dynamics}

We now turn to the long-time dynamics. To the degree that the IPL approximation holds over a range of state points, the long-time dynamics is a function of $\Gamma=\rho^{\gamma}/T$ where $\gamma$ is one third of the IPL exponent $n$, i.e., the ``density'' scaling of Eq. (1) is predicted to apply. This type of scaling has in recent years become a well established empirical fact for a large number of viscous liquids \cite{densscal,cas04} (usually ignoring the insignificant state point dependence of $t_0$). In particular, density scaling applies for van der Waals liquids, but not for liquids with hydrogen bonds or other directional bonds \cite{grz06,rol06,gra07,rol08}. Other forms of scaling have been suggested, but it has been argued that the scaling reflects an underlying IPL \cite{cas04,rol07,ped_prl,bai08,cos08,sch08}, and thus that Eq. (1) is the correct scaling.

A potential complication for the IPL explanation of density scaling, is that the IPL exponent suggested by the fluctuations ($n=3\gamma$) varies somewhat with state point \cite{nick} \footnote{Another issue is: what is the optimal exponent to use when correlation is not perfect? Any exponent $\gamma R^k$ ($k$ being a constant) has the required property of reducing to the pure IPL exponent in the limit of R going to one. In the present work we chose to test the exponent $\gamma$ defined in Fig.~\ref{Fig:Fluct}}. This is illustrated for the two models in Fig.~\ref{Fig:Slope}. Note that, while the suggested exponent is weakly state-point dependent for both models, one finds quite different exponents for the two models.
%To apply density scaling (Eq.~\ref{1}) a single value of $\gamma$ is needed.
For the OTP model we test density scaling  using the slope found by averaging over all state points: $\gamma =7.9$. For the dumbbell model we consider two values of the scaling exponent: the slope averaged over all state points,  $\gamma = 6.1$, and the slope averaged over the three data sets with the smallest slopes, 
%($\rho = 1.109$ g/cm$^3$, $\rho = 1.166$ g/cm$^3$ and P=2.0GPa); 
$\gamma =5.9$, i.e., the ``best'' compromise if the $\rho = 1.006$ g/cm$^3$ isochore is left out of consideration. 
%For the OTP model we will consider the slope averaged over all state points: $\gamma =7.9$.
%In the following we apply density scaling to the diffusion coefficient estimated from the long-time behavior of the mean-square displacement, $\left< \Delta r^2(t) \right>$, of the large spheres (the ``phenyl group'') \cite{Equil}. The diffusion coefficients for all state points studied are given in Fig.~\ref{Fig:D}. 
We apply density scaling to the reduced diffusion coefficient, $D^* \equiv (N/V)^{1/3}(k_BT/m)^{-1/2}D$ where $m$ is the mass of the molecules and $D$ is the diffusion coefficient estimated from the long-time mean-square displacement. %For the dumbbell model, the diffusion coefficients were estimated from the long-time behavior of the mean-square displacement, $\left< \Delta r^2(t) \right>$, of the large spheres (the ``phenyl group'') \cite{Equil}. For the OTP model the mean-square displacement of the central sphere was used. 

\begin{figure}
\begin{center}
\includegraphics[width=8.5cm]{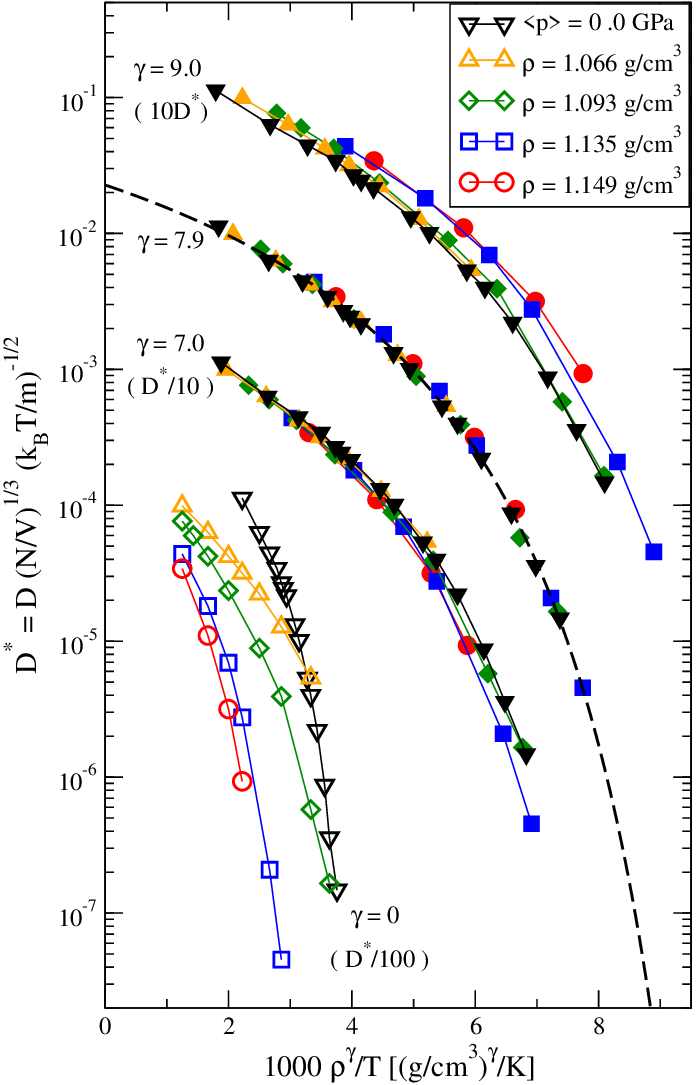}
 %\includegraphics[width=12.5cm]{fig5.eps}
 % FsB.eps: 1197x34 pixel, 0dpi, -3674644960619337680330868930316861440.00x-104375884446234152441730942875205632.00 cm, bb=
\end{center}
\caption{(Color online)
The reduced diffusion coefficient for Lewis-Wahnstr\"om OTP \cite{OTP}. 
Open symbols: $D^*$ plotted without scaling ($\gamma=0$, $D^*$ divided by 100). 
Filled symbols: $D^*$ scaled according to Eq.~(1) with three different scaling exponents:
$\gamma = 9.0$ (Upper set of curves, $D^*$ multiplied by 10), 
$\gamma = 7.9$ (Second set of curves), 
$\gamma = 7.0$ (Third set of curves, $D^*$ divided by 10).
%$\gamma = 0.0$ (Lower set of curves, Open symbols, $D^*$ divided by 100).
%As a guide to the eye, the expression 
%$D^* = 2.27\times 10^{-2} \exp\left( -353/(T/\rho^{7.9} - 87.8)\right)$ is plotted as a 
%fit to the data with  $\gamma = 7.9$. 
}
\label{Fig:DScaled_OTP}
\end{figure}

\begin{figure}
\begin{center}
 \includegraphics[width=8.5cm]{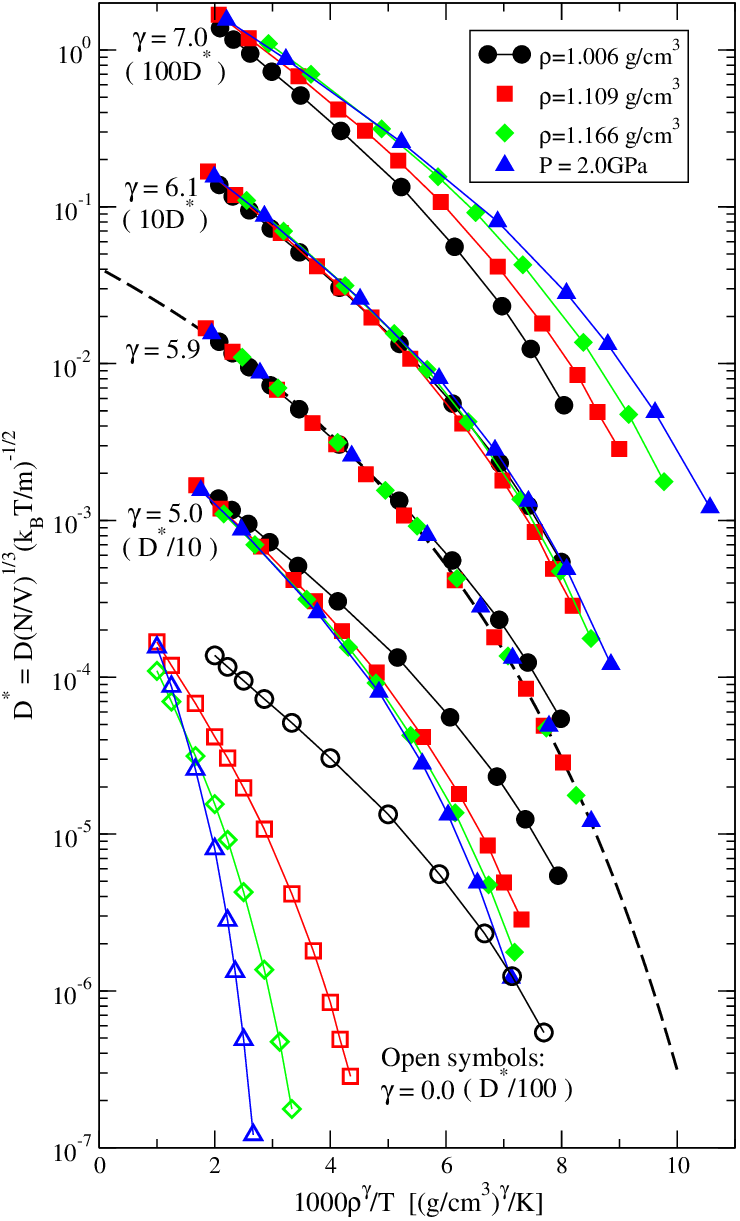}
 %\includegraphics[width=12.5cm]{fig6.eps}
 % FsB.eps: 1197x34 pixel, 0dpi, -3674644960619337680330868930316861440.00x-104375884446234152441730942875205632.00 cm, bb=
\end{center}
\caption{(Color online)
The reduced diffusion coefficient %,  $D^* \equiv (N/V)^{1/3}(k_BT/m)^{-1/2}D$, 
for the asymmetric dumbbell model \cite{Dumbbell}. 
Open symbols: $D^*$ plotted without scaling ($\gamma=0$, $D^*$ divided by 100). 
Filled symbols: $D^*$ scaled according to Eq.~(1) with four different scaling exponents:
$\gamma = 7.0$ (Upper set of curves, $D^*$ multiplied by 100), 
$\gamma = 6.1$ (Second set of curves, $D^*$ multiplied by 10), 
$\gamma = 5.9$ (Third set of curves), and 
$\gamma = 5.0$ (Fourth set of curves, $D^*$ divided by 10).
%As a guide to the eye, the expression 
%$D^* = 4.07\times 10^{-2} \exp\left( -462/(T/\rho^{5.9} - 60.8)\right)$ is plotted as a 
%fit to the three collapsing curves for $\gamma = 5.9$.
}\label{Fig:DScaled}
\end{figure}

Fig.~\ref{Fig:DScaled_OTP} applies density scaling to the Lewis-Wahnstr\"om OTP model. The scaling works neither with $\gamma=9.0$ nor with $\gamma=7.0$, whereas it works well with $\gamma=7.9$ which is the exponent we calculated from the fluctuations (Fig. \ref{Fig:Slope}). In Fig.~\ref{Fig:DScaled} density scaling is applied to the asymmetric dumbbell model. Density scaling works neither with $\gamma=7.0$ nor with $\gamma=5.0$. Comparing the scaling with $\gamma = 6.1$ to the data without scaling ($\gamma=0$, open symbols), we find quite good data collapse -- by far most of the density dependence is captured by using $\gamma = 6.1$.  With $\gamma = 5.9$, however, the data collapse is even better for three of the data sets, whereas one data set deviates from the master curve comprised of these three sets. The latter is the isochore $\rho=1.006$ g/cm$^3$, i.e., precisely the one that was ignored when choosing $\gamma = 5.9$ (Fig.~\ref{Fig:Slope}).

The above results demonstrate that density scaling is a consequence of an underlying IPL scale invariance which is revealed by studying thermal equilibrium fluctuations at a single state point. Obviously the  scaling with a single value of $\gamma$ can only be perfect for true IPL liquids. For ``real`` liquids a larger region of state points makes the scaling less perfect for two reasons (compare Fig.~\ref{Fig:DScaled}, $\gamma=6.1$ and $5.9$): The apparent IPL exponent ($n=3\gamma$) changes more, and the relative effect of intra-molecular interactions that do not obey the IPL scaling (here the fixed bonds) is larger. In this connection it is important to note, however, that even small density changes are experimentally relevant. Thus changing the density of, e.g., phenylphthalein-dimethylether by just 2\% can change the relaxation time by more than two ordes of magnitude \cite{PaluchPDE}.

\begin{figure}
\begin{center}
 \includegraphics[width=8.5cm]{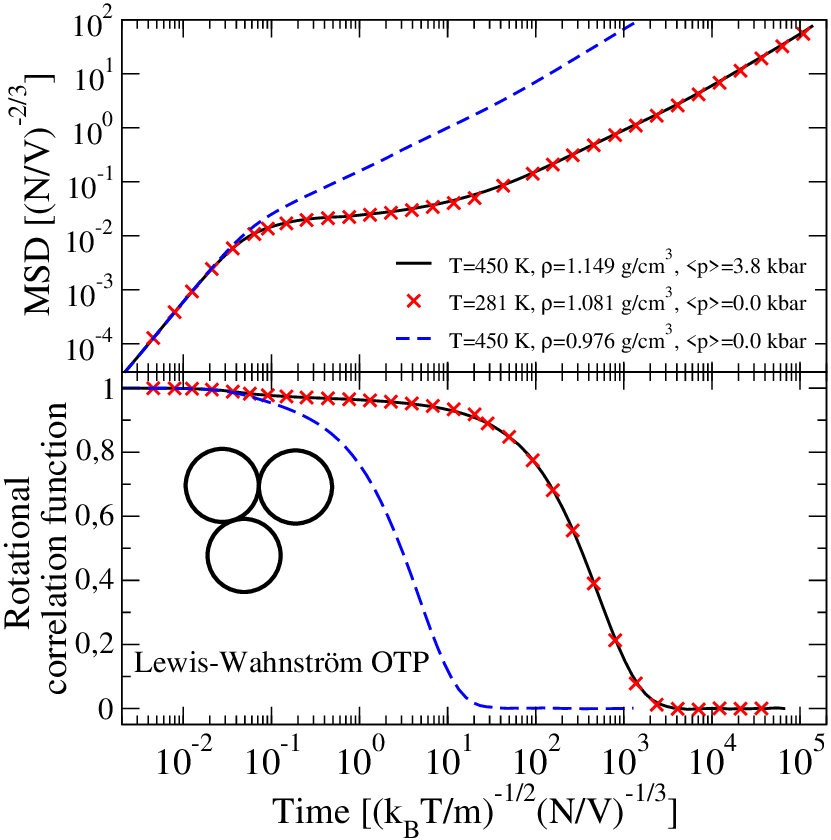}
 %\includegraphics[width=12.5cm]{fig7.eps}
 % FsB.eps: 1197x34 pixel, 0dpi, -3674644960619337680330868930316861440.00x-104375884446234152441730942875205632.00 cm, bb=
\end{center}
\caption{(Color online)
Dynamics of the LW OTP model \cite{OTP} in reduced units for the three state point shown in Fig.~\ref{Fig:grOTP}. {\bf a)} Mean -square displacement for the non-central LJ spheres. {\bf b)} Rotational correlation function for the vector perpendicular to the plane of the three LJ spheres, using the first Legendre polynomial.
}
\label{Fig:msd_OTP}
\end{figure}

As a consequence of the underlying IPL scale invariance, state points with same $\rho^\gamma/T$ are expected to have not only the same scaled structure and reduced diffusion coefficient as demonstrated above, every aspect of the dynamics should be the same in reduced units. Figure~\ref{Fig:msd_OTP} shows the reduced mean-square displacement and rotational correlation coefficient of the LW-OTP model, at the three state points of Fig.~\ref{Fig:grOTP}. The agreement between the two state points with same $\rho^\gamma/T$ (full black line and red crosses) is striking. In contrast, the third state point (blue dashed line) has the same temperature as the first state point (full black line) and the same pressure as the second (red crosses), but very different dynamics. Clearly, neither pressure nor temperature controls the dynamics -- to a very good approximation the controlling parameter is $\Gamma=\rho^{\gamma}/T$.

\begin{figure}
\begin{center} 
 \includegraphics[width=8.5cm]{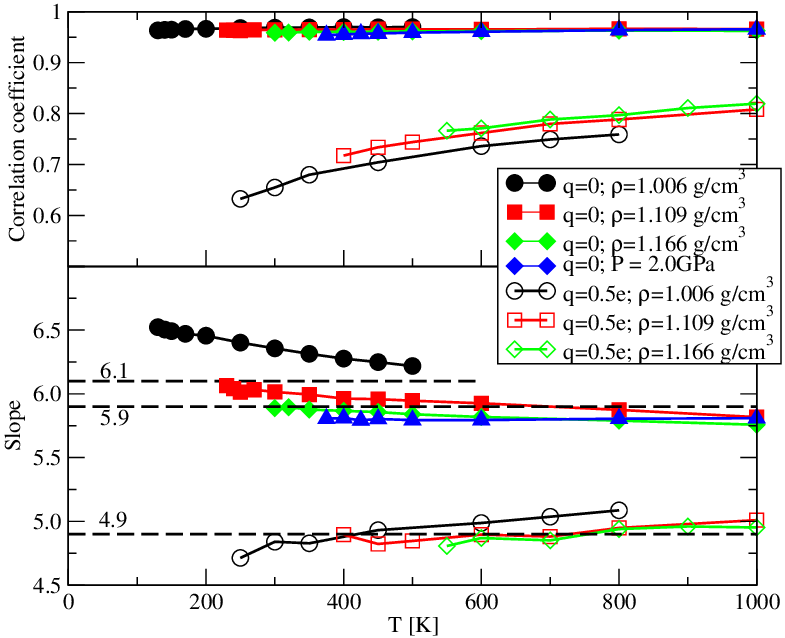}
\end{center}
\caption{(Color online)
Results from equilibrium molecular dynamics simulations of 512 asymmetric dumbbell molecules \cite{Dumbbell} with, respectively, a strong dipole moment ($q=\pm0.5$e, open symbols, three isochores), and zero dipole moment ($q=0$e, filled symbols, three isochores and an isobar). {\bf a)} Correlation coefficients, 
%describing the correlation of fluctuations in virial, $W$, and potential energy, $U$: 
$R\equiv \left< \Delta W \Delta U\right>/\sqrt{ \left< \Delta W^2 \right>\left< \Delta U^2 \right>}$. {\bf b)} The 'slopes', $\gamma \equiv \sqrt{\left< (\Delta W)^2 \right>/\left< (\Delta U)^2 \right>}$.
}\label{Fig:CCSlope}
\end{figure}

\begin{figure}
\begin{center}
 \includegraphics[width=8.5cm]{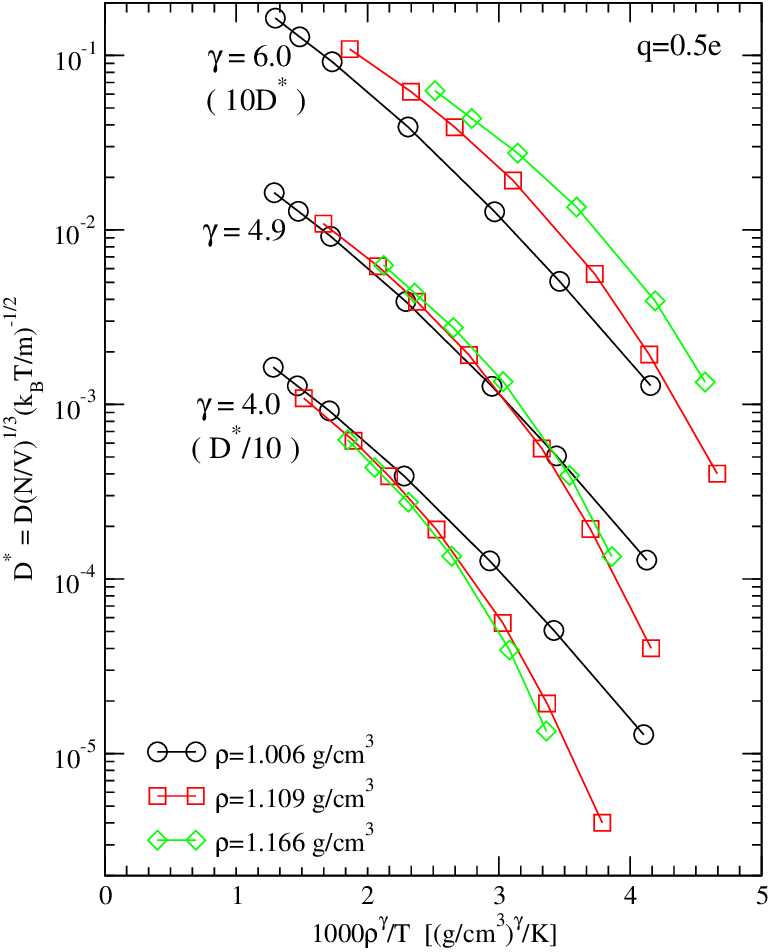}
\end{center}
\caption{(Color online)
The reduced diffusion coefficient, $D^* \equiv (N/V)^{1/3}(k_BT/m)^{-1/2}D$, for the asymmetric dumbbell model with $q=\pm0.5$e scaled according to Eq.~(1), with three different scaling exponents:
$\gamma = 6.0$ (Upper set of curves, $D^*$ multiplied by 10), 
$\gamma = 4.9$ (Middle set of curves), and 
$\gamma = 4.0$ (Lower set of curves, $D^*$ divided by 10).}
\label{Fig:DScaled_q}
\end{figure}

We argued above that the IPL scaling properties demonstrated is a feature only of strongly correlating liquids. To illustrate this we modified the asymmetric dumbbell model to make it less correlating. This was done by adding charges of $q=\pm0.5$e (e being the elementary charge) to the LJ spheres, resulting in a dipole moment of %0.29nm$\cdot$0.5e=2.32$\cdot10^{-29}$Cm= 
7.0D, i.e., %almost 20 times stronger than in toluene, %(0.36D) and 
almost 4 times stronger than water. The correlation coefficients and slopes are shown Fig. \ref{Fig:CCSlope} (open symbols), compared to the $q=0$ version of the model (filled symbols). The addition of the large dipole moments decreases the correlation, as expected. As before we use the averaged slope as the scaling exponent suggested by the fluctuations, giving $\gamma=4.9$ for $q=\pm0.5$e. The density scaling is applied in  Fig.~\ref{Fig:DScaled_q}. Density scaling with $\gamma=4.9$ works better than either $\gamma=6.0$ or $\gamma=4.0$, but the data collapse is clearly inferior to what is seen in Fig.~\ref{Fig:DScaled} and Fig.~\ref{Fig:DScaled_OTP} (note the smaller dynamic range in Fig.~\ref{Fig:DScaled_q}). This confirms that density scaling is indeed a property of strongly correlating liquids, consistent with the experimental finding that, e.g., hydrogen-bonded liquids do not obey density scaling.  

\section{Conclusions}

We have provided direct evidence for an underlying IPL scale invariance in model molecular van der Waals liquids, leading to three experimentally testable predictions for these liquids: i) The density scaling exponent $\gamma$ can be identified from the equilibrium fluctuations via the fluctuation-dissipation theorem determining the frequency-dependent linear thermoviscoelastic response functions \cite{ped_pre,ell07}; ii) different state points with same $\rho^\gamma/T$ -- same values of $D^*$ and $\tau_\alpha$ -- are ''isomorphic'', i.e., have same scaled structure \cite{gnan09}. iii) Isomorphic state points must have the same relaxation spectra. This is consistent with recent experimental findings for molecular glass formers and amorphous polymers (excluding hydrogen bonds): 
Varying pressure and temperature, state points with the same relaxation time have the same shape of the dielectric $\alpha$-relaxation function \cite{ngai05,roland08}. Finally, the hidden scale invariance explanation of density scaling explains why one observes density scaling of the dynamics, without the same scaling applying to the pressure \cite{densscal,grz09}.
%for each liquid, the shape of the dielectric $\alpha$-relaxation function only depends on the relaxation time \cite{ngai05}.

The hidden scale invariance is a consequence of the strong virial/potential energy correlations \cite{ped_prl,nick}  characterizing van der Waals liquids, as well as most or all metallic liquids. Thus the results from computer simulations of two model liquids presented here point to a new physics of these large classes of liquids, which is simpler than previously believed.

\acknowledgments 
%{\bf Acknowledgments:}    
Useful discussions with Kristine Niss are gratefully acknowledged.
This work was supported by the Danish National Research Foundation's (DNRF) centre for viscous liquid dynamics ``Glass and Time.''


\begin{thebibliography}{99}
%(REF: Hoover, Ross, evt. Ben-Amotz 2003; Japaner paper fra 1980'erne om dynamikken)
\bibitem{IPL}
W. G. Hoover, M. Ross, K. W. Johnson, D. Henderson, J. A. Barker,  and B. C. Brown, J. Chem. Phys. {\bf 52}, 4931 (1970);
W. G. Hoover, S. G. Gray, and K. W. Johnson, J. Chem. Phys. {\bf 55}, 1128 (1971);
Y. Hiwatari, H. Matsuda, T. Ogawa, N. Ogita, and A. Ueda, Prog. Theor. Phys. {\bf 52}, 1105 (1974);
D. Ben-Amotz and G. J. Stell, J. Chem. Phys. {\bf 119}, 10777 (2003);
C. De Michele, F. Sciortino, and A. Coniglio, J. Phys.: Condens. Matter {\bf 16},  L489 (2004).

\bibitem{ped_pre} U. R. Pedersen, T. Christensen, T. B. Schr{\o}der, and J. C. Dyre, Phys. Rev. E {\bf 77}, 011201 (2008).

\bibitem{ped_prl}  U. R. Pedersen, N. P. Bailey, T. B. Schr{\o}der, and J. C. Dyre, Phys. Rev. Lett. {\bf 100}, 015701 (2008).

\bibitem{nick} N. P. Bailey,  U. R. Pedersen, N. Gnan, T. B. Schr{\o}der, and J. C. Dyre, J. Chem. Phys. {\bf 129}, 184507 (2008); J. Chem. Phys. {\bf 129}, 184508 (2008).

\bibitem{SimDet}
NVT simulations were carried out using Gromacs software  \cite{gromacs} using the Nos\'{e}-Hoover thermostat \cite{NoseHoover}. Molecules were kept rigid using the LINCS \cite{LINCS} algorithm.

\bibitem{gromacs} H. J. C. Berendsen, D. van der Spoel, and R. van Drunen, { Comp. Phys. Comm.} {\bf 91}, 43 (1995);  E. Lindahl, B. Hess, and D. van der Spoel, {J. Mol. Mod.} {\bf 7}, 306 (2001).

\bibitem{NoseHoover} S. A. Nos\'e, { Mol. Phys.} {\bf 52}, 255 (1984); W. G. Hoover, {Phys. Rev. A} {\bf 31}, 1695 (1985).

\bibitem{LINCS} B. Hess, H. Bekker, H. J. C. Berendsen, and J. G. E. M. Fraaije, J. Comp. Chem. {\bf18}, 1463 (1997). %OK-ulf

\bibitem{Dumbbell} 
The large LJ sphere, mimicking the phenyl group, is similar to the one in the Lewis-Wahnstr\"om OTP model \cite{wahn94} with the parameters $m_p=77.106$ u, $\sigma_p=0.4963$ nm and $\epsilon_p=5.726$ kJ/mol. The small sphere, mimicking the methyl group, was taken from UA-OPLS \cite{Jorg84} with $m_m=15.035$ u, $\sigma_m=0.3910$ nm and $\epsilon_m=0.66944$ kJ/mol. 
Bond length: $d=0.29$ nm. The interaction between unlike particles is determined by the Lorentz-Berthelot mixing rules. 

\bibitem{wahn94} %J. L. Laurent and G. Wahnstr\"om, Phys Rev. E {\bf50}, 3865 (1994). %OK-ulf
G. Wahnstr\"om and L. J. Lewis, Physica A {\bf 201}, 150 (1993); L. J. Lewis and G. Wahnstr\"om, Phys. Rev. E {\bf 50}, 3865 (1994).

\bibitem{Jorg84} W. L. Jorgensen, J. D. Madura, and C. J. Swenson, J. Am. Chem. Soc. {\bf106}, 6638 (1984). %OK-ulf


\bibitem{OTP}
$N=324$ molecules consisting of three LJ particles (with $\sigma=0.483$ nm, $\varepsilon= 600 $ K$k_B \simeq 4.989$ kJ/mol and $m=76.768$ u) placed in the corners of a rigid isosceles triangle with two sides of length $\sigma=0.483$ nm and one angle of $75^{\circ}$ \cite{wahn94}. LJ potentials were cut at $r_c=2.5\sigma$ using a shift function \cite{gromacs} with $r_1=2.3\sigma$. 

\bibitem{densscal} 
A. T{\"o}lle, Rep. Prog. Phys. {\bf 64}, 1473 (2001); %OK 
C. Dreyfus, A. Aouadi, J. Gapinski, M. Matos-Lopes, W. Steffen, A. Patkowski, and R. M. Pick, Phys. Rev. E {\bf 68}, 011204 (2003); %OK
C. Alba-Simionesco, A. Cailliaux, A. Alegria, and G. Tarjus, Europhys. Lett. {\bf 68}, 58 (2004); 
C. M. Roland, S. Hensel-Bielowka, M. Paluch, and R. Casalini, Rep. Prog. Phys. {\bf 68}, 1405 (2005).

\bibitem{cas04} R. Casalini and C. M. Roland, Phys. Rev. E {\bf69}, 062501 (2004).
\bibitem{grz06} A. Grzybowski, K. Grzybowska, J. Ziolo, and M. Paluch, Phys. Rev E {\bf74}, 041503 ͑(2006͒).
\bibitem{rol06} C. M. Roland, S. Bair, and R. Casalini, J. Chem. Phys. {\bf125}, 124508 ͑(2006͒).
\bibitem{gra07} A. Le Grand, C. Dreyfus, C. Bousquet, and R. M. Pick, Phys. Rev E {\bf75}, 061203 ͑(2007͒).
\bibitem{rol08} C. M. Roland, R. Casalini, R. Bergman, and J. Mattsson, Phys. Rev. B {\bf 77}, 012201 (2008).

\bibitem{rol07} C. M. Roland and R. Casalini, J. Phys.: Condens. Matter {\bf 19},  205118 (2007).
\bibitem{bai08} N. P. Bailey, T. Christensen, B. Jakobsen, K. Niss,
N. B. Olsen, U. R. Pedersen, T. B. Schr{\o}der, and J. C. Dyre, J. Phys.: Condens. Matter {\bf 20} 244113 (2008).

\bibitem{cos08} D. Coslovich and C. M. Roland, J. Phys. Chem. B  {\bf 112}, 1329 (2008);
 J. Chem. Phys. {\bf 130}, 014508 (2009).

\bibitem{sch08} T. B. Schr\o{}der, U. R. Pedersen, and J. C. Dyre, arXiv:0803.2199 (2008).

\bibitem{PaluchPDE} M. Paluch, R. Casalini, and C. M. Roland, Phys. Rev. B {\bf 66}, 092202 (2002).

\bibitem{ell07} 
N. L. Ellegaard, T. Christensen, P. V, Christiansen, N. B. Olsen, U. R. Pedersen, T. B. Schr{\o}der, and J. C. Dyre, J. Chem. Phys. {\bf 126}, 074502 (2007).

\bibitem{gnan09} N. Gnan, T. B. Schr\o{}der, U. R. Pedersen, N. P. Bailey, and J. C. Dyre, arXiv:0905.3497 (2009).

\bibitem{ngai05}  K. L. Ngai, R. Casalini, S. Capaccioli, M. Paluch, and C. M. Roland,  J. Phys. Chem. B  {\bf 109}, 17356 (2005).
\bibitem{roland08}  C. M. Roland,  Soft Matter  {\bf 4}, 2316 (2008).

\bibitem{grz09} A. Grzybowski, M. Paluch, and K. Grzybowska, J. Phys. Chem. B {\bf 113},  7419 (2009).

\end{thebibliography}
\end{document}